\documentclass[apjl,usenatbib]{emulateapjFLF}

\usepackage{amsmath}
\usepackage{natbib}
\usepackage{color}
\usepackage{graphicx}
\usepackage{xspace}
\usepackage[normalem]{ulem}
\usepackage{multirow}
\usepackage{comment}

\newcommand{\MBH}{M$_{\rm BH}$}
\newcommand{\sig}{$\sigma^2_{\rm rms}$}

\definecolor{orange}{rgb}{1,0.5,0}

\let\oldmarginpar\marginpar
\renewcommand\marginpar[1]{\-\oldmarginpar[\raggedleft\footnotesize #1]%
{\raggedright\footnotesize #1}}

\shorttitle{A new cosmological distance measure using AGN X-ray variability}
\shortauthors{La Franca et al.}

\begin{document}

\title{A new cosmological distance measure using AGN X-ray variability}
\author{Fabio La Franca\altaffilmark{1,2}, Stefano Bianchi\altaffilmark{1}, Gabriele Ponti\altaffilmark{3}, Enzo Branchini\altaffilmark{1,2,4}, and
Giorgio Matt\altaffilmark{1}}

\altaffiltext{1}{Dipartimento di Matematica e  Fisica, Universit\`a Roma Tre,
via della Vasca Navale 84, I-00146, Roma, Italy; lafranca,bianchi,branchin,matt@fis.uniroma3.it}

\altaffiltext{2}{INAF - Osservatorio Astronomico di Roma, via Frascati 33, I-00040, Monte Porzio Catone (RM), Italy}

\altaffiltext{3}{Max-Planck-Institut f\" ur extraterrestrische Physik, Giessenbachstrasse 1, D-85748 Garching bei M\" unchen, Germany}

\altaffiltext{4}{INFN - Sezione di Roma Tre, via della Vasca Navale 84, I-00146, Roma, Italy}

\begin{abstract}
We report the discovery of a luminosity distance estimator  using Active Galactic Nuclei (AGN).
We combine the correlation between the X-ray variability amplitude and the Black Hole (BH) mass
with the single epoch spectra BH mass estimates which depend on the AGN luminosity and the line width emitted by the 
broad line region. We demonstrate that significant correlations do exist which allows one to predict the AGN (optical or X-ray) luminosity
as a function of the AGN X-ray variability and either the H$\beta$ or the Pa$\beta$ line widths. In the best case, when the Pa$\beta$ is used, the relationship has an intrinsic dispersion
of $\sim$0.6 dex. Although intrinsically more disperse than Supernovae Ia, this relation constitutes an alternative distance indicator potentially able to probe, in an independent way, the expansion history of the Universe. With this respect, we show that the new mission concept {\it Athena}
should be able to measure the X-ray variability of hundreds of AGN and then constrain the distance modulus with uncertainties of 0.1 mag up to $z \sim 0.6$. We also discuss how, using a new dedicated wide field X-ray telescope able to measure the variability of thousands
of AGNs, our estimator has the prospect  to become a cosmological probe even more sensitive than current Supernovae Ia samples.

\end{abstract}


\maketitle

\section{Introduction}
\setcounter{footnote}{0}

One of the most important results on observational cosmology is the
discovery, using type Ia supernovae (SNeIa) as standard candles, of the accelerating
expansion of the Universe \citep{riess98, perlmutter99}. However, the use of  SNeIa
is  difficult beyond $z\sim 1$ and limited up to $z\sim2$ \citep[e.g.][]{rubin13}. It is therefore of paramount importance to calibrate
other independent distance indicators able to measure the Universe expansion. It would be even better if such a method would be able to probe even beyond these redshifts, where the differences among various cosmological models are larger.

Given their high luminosities, since their discovery there have been several studies  on the use of Active Galactic Nuclei (AGN) as standard candles or rulers
\citep{baldwin77, rudge99,collier99,  elvis02}. More recently, also thanks to a better understanding
of the AGN structure, more promising methods have been presented  (see \citet{marziani13} and the review therein). For example, many authors  \citep[e.g.][]{watson11} 
use the tight relationship between the luminosity of an AGN and the radius of its Broad Line Region (BLR) established via reverberation mapping to determine the luminosity distances. On the other hand, \citet{wang13} suggest that super-Eddington accreting massive BH may reach saturated luminosities, which then  provide a new tool for estimating cosmological distances.
Besides AGN, gamma ray bursts  (GRB) have been used as standard candles,  however their low identification rate 
makes their use difficult \citep[e.g.][and references therein]{schaefer07}.

Here we propose a new method to predict the AGN luminosity based on the combination of 
the virial relations, which allow to derive the  BH mass (M$_{\rm BH}$) from the AGN luminosity and the width of the lines emitted from  the BLR, and the well established anti-correlation between \MBH\ and the  X-ray variability amplitude.

\section{Method}

\begin{deluxetable*}{l c c c c c  c c c c }
\tablecaption{Calibration Sample}
\tablewidth{0pt}
\tablehead{
\colhead{Name} & \colhead{z} & \colhead{$\sigma^2_{\rm rms}$} & \colhead{low-err$_{\sigma^2_{\rm rms}}$} & \colhead{up-err$_{    \sigma^2_{\rm rms}   }$} & \colhead{logL$_{\mathrm{5100}}$} & \colhead{logL$_{\mathrm{x}}$} & \colhead{FWHM$_{H\beta}$} & \colhead{FWHM$_{Pa\beta}$} & \colhead{Ref}\\
\colhead{} & \colhead{} & \colhead{} & \colhead{} & \colhead{}  & \colhead{\scriptsize erg  s$^{-1}$} & \colhead{\scriptsize erg s$^{-1}$} & \colhead{km s$^{-1}$} & \colhead{km s$^{-1}$} & \colhead{ }\\
\colhead{(1)} & \colhead{(2)} &\colhead{(3)} &\colhead{(4)} &\colhead{(5)} &\colhead{(6)} &\colhead{(7)} &\colhead{(8)} &\colhead{(9)} & \colhead{(10)}
}
\startdata
1E0919+515              & 0.1610 & 0.078000 & 0.052439 & 0.091463 & 44.29  & 43.43 &  1980 &     ... & a \\
1H0707-495              & 0.0411 & 0.219000 & 0.027439 & 0.030488 &     ...  & 42.67 &  1000 &     ... & b\\  
3C120                   & 0.0330 & 0.000210 & 0.000116 & 0.000183 & 44.09  & 44.06 &  2327 &  2733  & c \\   
3C273                   & 0.1583 & 0.000027 & 0.000022 & 0.000024 & 46.02  & 45.80 &  3500 &  2916 & d  \\   
ARK120                  & 0.0323 & 0.000290 & 0.000249 & 0.000427 & 44.37  & 43.96 &  6120 &  5114 & d \\   
ARK564                  & 0.0247 & 0.044000 & 0.008537 & 0.012195 &     ...  & 43.37 &     ... &  1616 &  \\
ESO198-G24              & 0.0455 & 0.000780 & 0.000445 & 0.000671 &     ...  & 43.70 &  6400 &     ... & e\\   
HE1029-1401             & 0.0860 & 0.001000 & 0.000610 & 0.001159 & 45.27  & 44.54 &  7500 &     ...  & f\\
HE1143-1810             & 0.0330 & 0.000690 & 0.000463 & 0.000671 &     ...  & 43.85 &  2400 &     ...  & e \\ 
IC4329A                 & 0.0161 & 0.000168 & 0.000084 & 0.000137 & 42.89  & 43.75 &  5964 &     ... & c\\   
IRAS13349+2438          & 0.1076 & 0.007300 & 0.002256 & 0.004268 & 44.64  & 43.87 &  2800 &     ...&  g \\   
IRAS17020+4544          & 0.0610 & 0.016400 & 0.004695 & 0.007317 &     ...  & 43.71 &  1040 &     ...  & h \\ 
IRASF12397+3333         & 0.0450 & 0.009500 & 0.003049 & 0.004878 & 43.36  & 43.38 &  1640 &     ... & g \\   
MCG-6-30-15             & 0.0077 & 0.035200 & 0.005000 & 0.004878 & 42.86  & 42.75 &  2020 &     ... & i \\
MRK110                  & 0.0353 & 0.000398 & 0.000269 & 0.000410 & 43.63  & 43.91 &  2079 &  1918  & d\\   
MRK1502                 & 0.0611 & 0.021300 & 0.005732 & 0.012195 & 44.79  & 43.65 &  1171 &     ...  & d\\ 
MRK279                  & 0.0305 & 0.000275 & 0.000173 & 0.000238 & 43.82  & 43.78 &  5411 &  3568  & d\\   
MRK335                  & 0.0258 & 0.015160 & 0.002382 & 0.002382 & 43.71  & 43.44 &  1841 &  1858  & d\\   
MRK509                  & 0.0344 & 0.000296 & 0.000093 & 0.000091 & 43.91  & 44.02 &  3424 &  3077  & l\\   
MRK590                  & 0.0264 & 0.002154 & 0.001468 & 0.002390 & 44.01  & 43.04 &  2627 &  3964  & d\\   
MRK766                  & 0.0129 & 0.027787 & 0.003285 & 0.003285 & 43.31  & 42.94 &  1100 &     ...  & g\\ 
MRK841                  & 0.0364 & 0.001700 & 0.001037 & 0.001829 & 43.64  & 43.49 &  6000 &     ...  & g\\   
NGC3227                 & 0.0039 & 0.008757 & 0.001976 & 0.004366 & 42.86  & 41.57 &  4445 &  2955  & g\\   
NGC3516                 & 0.0088 & 0.003594 & 0.001015 & 0.001015 & 43.17  & 42.46 &  5236 &  4469  & g\\  
NGC3783                 & 0.0097 & 0.005052 & 0.001217 & 0.001217 & 43.20  & 43.08 &  3555 &     ...  & d\\ 
NGC4051                 & 0.0023 & 0.113421 & 0.012172 & 0.012172 & 41.47  & 41.44 &  1170 &  1681  & l\\  
NGC4151                 & 0.0033 & 0.000890 & 0.000427 & 0.000915 & 42.58  & 42.53 &  6421 &  4667  & d\\   
NGC4395                 & 0.0011 & 0.144050 & 0.031327 & 0.065367 &     ...  & 40.21 &  1500 &     ... &   m\\
NGC4593                 & 0.0090 & 0.009819 & 0.003163 & 0.006379 & 42.85  & 42.87 &  5143 &  3791 &  c \\   
NGC5548                 & 0.0172 & 0.000283 & 0.000173 & 0.000278 & 43.21  & 43.42 &  6300 &  6525  & l\\   
NGC7469                 & 0.0163 & 0.002942 & 0.000641 & 0.001363 & 43.74  & 43.23 &  2639 &  1792  & d\\   
PDS456                  & 0.1840 & 0.004800 & 0.001402 & 0.001220 &     ...  & 44.90 &  3974 &  2068 & n \\
PG0844+349              & 0.0640 & 0.019000 & 0.010366 & 0.024390 &     ...  & 43.70 &     ... &  2410 \\   
PG1211+143              & 0.0809 & 0.010500 & 0.002683 & 0.005488 & 44.42  & 43.73 &  1900 &     ... & g\\   
PG1440+356              & 0.0791 & 0.006400 & 0.002927 & 0.004878 & 44.46  & 43.61 &  1630 &     ...  & g\\   
RE1034+396              & 0.0421 & 0.018000 & 0.006707 & 0.009756 & 43.18  & 42.53 &   ~~700 &     ... & g\\   
RXJ0057.2-2223          & 0.0620 & 0.023000 & 0.010366 & 0.018293 & 44.27  & 43.65 &   ~~970 &     ...  & g\\   
RXJ0136.9-3510          & 0.2890 & 0.064000 & 0.026220 & 0.054878 &     ...  & 44.34 &  1320 &     ... & o  \\  
RXJ0323.2-4931          & 0.0710 & 0.009000 & 0.006707 & 0.010976 & 43.58  & 43.21 &  1680 &     ...  & g\\  
SDSSJ135724.51+652505.9 & 0.1063 & 0.073000 & 0.044512 & 0.054878 & 43.14  & 42.90 &   ~~737 &     ... & p    
\enddata
\tablecomments{ Columns 3, 4 and 5: Excess variance, \sig,  and lower and upper errors at 1$\sigma$ confidence level from \citet{ponti12}. Column 6: $\lambda L_\lambda$  continuum luminosity at 5100\AA . Column 7: 2-10 keV intrinsic luminosity. Column 9: from \citet{landt08, landt13}. Column 10:
References for FWHM$_{H\beta}$ and L$_{\mathrm{5100}}$:
 a) \citet{jin12}, b) \citet{boller96}, c) \citet{assef12}, d) \citet{vestergaard06}, e) \citet{winkler92}, f) \citet{mclure01}, g) \citet{grupe04}, h) \citet{leighly99}, i) \citet{mchardy05}, l)  \citet{wandel99}, m) \citet{kraemer99},  n) \citet{torres97}, o) \citet{grupe99}, p) \citet{zhou06}.}
\label{tab:sample}
\end{deluxetable*}

The method which uses single epoch (SE) spectra (in the optical or near infrared bands) to measure \MBH\ 
\citep[e.g.][]{wandel99,  mclure02, vestergaard09, landt13}  is now well established.
By combining the velocity, $\Delta V$, of the BLR clouds (assuming Keplerian orbits) along with their distance R it is possible to determine the total mass contained within the BLR (which is dominated by the BH) using
\begin{equation}
{\rm M}_{\rm BH}= {{f \Delta {\rm V}^2 {\rm R}}\over{\rm G}},
\end{equation}

\noindent
where G is the gravitational constant, $\Delta {\rm V}^2 {\rm R}$ is the virial product and {\it f} is a factor which depends on the geometric and kinematic structure of the BLR.
These techniques derive \MBH\ using SE spectra to measure $\Delta$V from the Full Width at Half Maximum (FWHM) of 
some of the BLR lines (typically: H$_\beta$ or MgII2798\AA\ or CIV1459\AA ) and R from either the continuum or the line luminosities, L, which have been proved to be proportional to R$^2$ \citep[see][and references therein]{bentz13}. Therefore, the SE estimates are based on relations of the type

\begin{equation}
{\rm log M}_{\rm BH}  = \alpha  {\rm log L}+ \beta  {\rm log} \Delta V + \gamma, 
\label{eq:mrev}
\end{equation}

\noindent
where the values of the parameters
$\alpha\sim 0.5$, $\beta\sim 2$ and $\gamma$  depend on the emission line of the BLR used.
These relationships have typical uncertainties  of $\sim$0.5 dex.

On the other hand, several studies have found a significant anti-correlation between M$_{\rm BH}$ and the X-rays variability \citep{nandra97,turner99,oneill05,mchardy06,gierlinski08,zhou10,ponti12,kelly13}. Following \citet{ponti12} it results

\begin{equation}
{\rm log M}_{\rm BH} = - k \cdot  {\rm log}\sigma_{\rm rms}^2 + w,
\label{eq:mvar}
\end{equation}

\noindent
where $k\sim 0.8$ and  $w$ are two constants which depend on the time scale and the energy range where the variable flux is measured.
$\sigma^2_{rms}$ is the normalised excess variance
variability  estimator:

\begin{equation}
\sigma^2_{\rm rms}= {1\over{N\mu^2}} \Sigma_{i=1}^N \left[(X_i-\mu)^2-\sigma_i^2  \right],
\label{eq:var}
\end{equation}

\noindent 
where N is the number of time intervals where the fluxes (or the counts) $X_i$  are measured, $\mu$
is the mean of the N fluxes $X_i$ and $\sigma_i$ is the uncertainty on the i-th flux measure.
According to X-ray variability studies on samples of AGN whose M$_{\rm BH}$ has been measured with reverberation
mapping techniques, these kinds of relationships could have spreads as narrow as 0.2-0.4 dex \citep{zhou10,ponti12,kelly13}.

The origin of \MBH\ -\sig relation (eq. \ref{eq:mvar}) is to be found in the dependence on \MBH\ of the time-scales of what appears to be a universal shape of the
AGN variability power spectral density (PSD). Indeed, the AGN X-ray PSDs are generally well modelled by 
two power laws, ${\rm P}(\nu)\propto 1/\nu^n $, where the PSD slope is $n \sim 1$ down to a 
break frequency $\nu_b$, that scales primarily with \MBH, and then steepens to $n\sim 2$ at larger frequencies.
However, measuring the shape of the AGN X-ray PSD is very data demanding, requiring high quality data on many different time-scales. Therefore these studies are confined to a relatively small number of sources. 
On the contrary, the excess variance $\sigma^2_{\rm rms}$ is a robust estimator as it corresponds to the
integral of the PSD on the time scales probed by the data. 
The scaling of the characteristic frequencies of the PSD with \MBH\ (and the roughly similar PSD normalisation at $\nu_b$) induces a dependence of the excess variance with \MBH\ (if computed at frequencies above $\nu_b$).

The two equations (\ref{eq:mrev} and \ref{eq:mvar}) used to estimate \MBH\ can be combined to
derive the intrinsic AGN luminosity as a function of its X-ray variability, $\sigma_{rms}^2$,
and line width, $\Delta V$:

\begin{equation}
{\rm log L} =  -{k\over \alpha} \cdot {\rm log} \sigma^2 - {\beta \over \alpha} {\rm log} \Delta V + {\rm const},
\label{Lcosmo1}
\end{equation}

\noindent
which, if we (for the sake of simplicity) assume $\alpha=0.5$ and $\beta=2$, becomes

\begin{equation}
{\rm log L} =  -2k \cdot {\rm log} \sigma^2 - 4 {\rm log} \Delta V + {\rm const}.
\label{eq:relation}
\end{equation}

In this work we aim to  verify if the above proposed relationship does work and then calibrate it.
It should be noted that, in many previous studies, a correlation between the AGN luminosity and X-ray variability has been measured \citep[e.g.][and references therein]{ponti12,shemmer14}.
According to the above discussion, we believe that such a correlation is the projection on the L-$\sigma^2_{\rm rms}$ plane of our proposed 3D relationship among L, $\sigma^2_{\rm rms}$ and $\Delta V$ (eq. \ref{eq:relation}).
If this is the case, we should measure a more significant and less scattered relation than previously reported using only
L and $\sigma^2_{\rm rms}$.

We adopted a flat cosmology with H$_0$ = 70 km s$^{-1}$ Mpc$^{-1}$, $\Omega_{\rm M}$=0.30 and $\Omega_\Lambda$=0.70. Unless otherwise stated, uncertainties are quoted at the 68\% (1$\sigma$) confidence level.

\section{Calibration Sample}

We have used the variability measures
coming from the  XMM-Newton systematic excess variance study of radio quiet, X-ray un-obscured, AGN by \citet{ponti12}.
Light curves have been constructed in the 2-10 keV energy band with t$_0$=250 s long bins and divided into segments of 20 ks.
We selected all those objects whose excess variance, $\sigma^2_{\rm rms}$, was measured with a significance larger than $\sim$1.2$\sigma$ and for which
the FWHM of the broad component of the
H$\beta$ (FWHM$_{{\rm H}\beta}$) and the $\lambda L_\lambda$ continuum luminosity at  5100\AA , L$_{\rm 5100}$, estimates were available in the literature. For most of the
objects, when possible, coeval measures, obtained from the same optical spectra, were used. In addition, we collected the Pa$\beta$
FWHM measures (FWHM$_{{\rm Pa}\beta}$) from \citet{landt08, landt13}, when available. In total the sample contains 40 low redshift AGN (86\% with $z< 0.1$),
38 and 18 with H$\beta$ or Pa$\beta$ line widths measurements available, respectively (in two objects, Ark 564 and PG 0844+349,
the H$\beta$ line width and L$_{\rm 5100}$ measures are missing, while the Pa$\beta$ line width is available; see Table \ref{tab:sample}).

\section{The calibration fits}

We have performed a linear fit of the relation

\begin{equation}
{\rm log}{\rm L\over{\rm erg~ s^{-1}}}  + 4 {\rm log}{{\rm FWHM}\over {\rm 10^3~ km~ s^{-1} }}  =  \alpha \cdot {\rm log} \sigma^2_{\rm rms}  + \beta,
\label{eq:relation2}
\end{equation}

\begin{figure*}
 \centering
  \includegraphics[height=15cm, angle=-90]{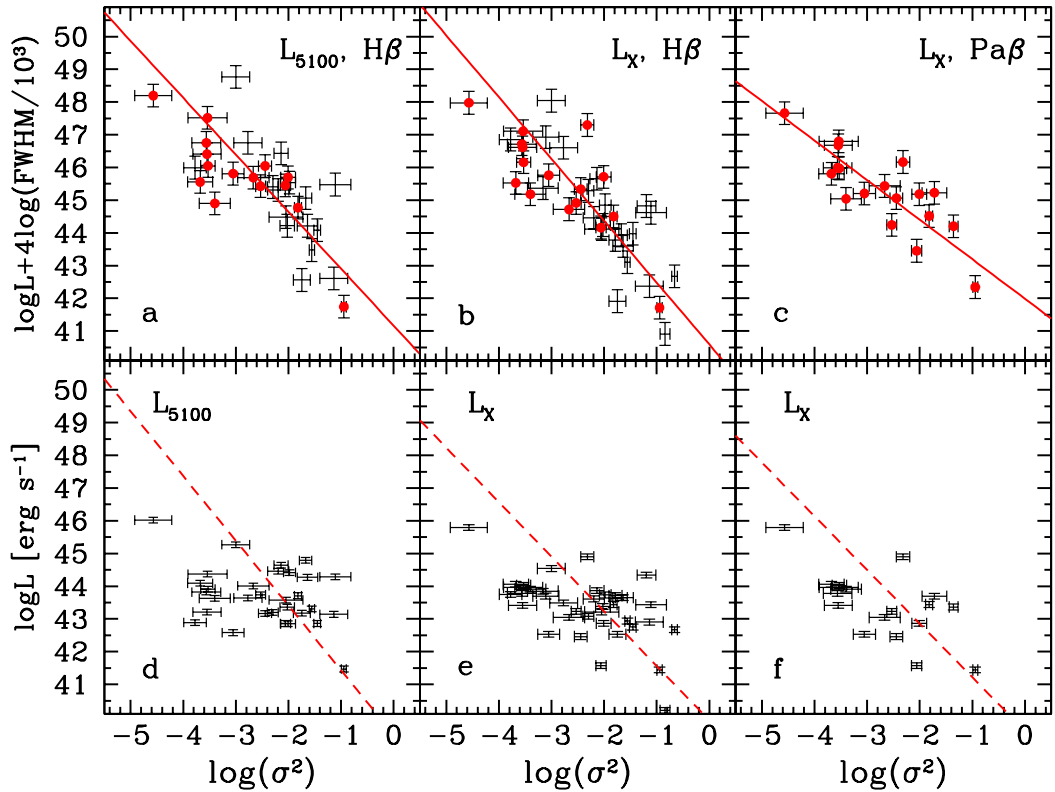}
  \caption{ (a) Square of the virial product (computed using L$_{\rm 5100}$ and FWHM$_{\rm H\beta}$)  as a function of the excess variance, \sig,  of the 2-10 keV flux measured into 20 ks long bins. The continuous lines show the fit to the data (see Table \ref{tab:fit}). The objects having the Pa$\beta$ FWHM measures available are shown by red filled circles. (b) As in the previous panel using L$_{\rm X}$ and FWHM$_{\rm H\beta}$. (c) As in the previous panel using L$_{\rm X}$ and FWHM$_{\rm Pa\beta}$. (d) L$_{\rm 5100}$ as a function of \sig\ of the same sample shown in the upper panel. (e) L$_{\rm X}$ as a function of \sig\ of the same sample shown in the upper panel. (f)  L$_{\rm X}$ as a function of \sig\ of the same sample shown in the upper panel. The dashed lines show the best fits to the data  (see Table \ref{tab:fit}).}
        \label{fig_1}
 \end{figure*}

\noindent
(looking for the best fit $\alpha$ and $\beta$ parameters) using FITEXY \citep{press07}  that can incorporate errors on both variables\footnote{We preferred to look for the dependence of the square of the virial product from \sig , instead of using eq. \ref{eq:relation}, in order to have comparable uncertainties on both axes (see Fig. \ref{fig_1}).}.
As a first step we have investigated whether a relationship exists using L$_{\rm 5100}$ and FWHM$_{{\rm H}\beta}$ to build up the virial product.
Among the three quantities $\sigma^2_{\rm rms}$, L$_{\rm 5100}$ and FWHM$_{{\rm H}\beta}$, we can ignore the uncertainties on L$_{\rm 5100}$ as they are experimentally less than 3\%, while the uncertainties on \sig\ are in the  range 10-90\% (see Table \ref{tab:sample} and references therein). As far as the FWHM measures are concerned, and although in some cases  the uncertainties are reported in the literature, we have preferred to assume a common uncertainty of 20\%
following the results of \citet{grupe04, vestergaard06, denney09a, assef12}.  The data and the results of the fit are shown in Figure \ref{fig_1} and Table \ref{tab:fit}. 
The logarithm of the square of the virial product, computed
using L$_{\rm 5100}$ and FWHM$_{{\rm H}\beta}$, is strongly correlated  with the logarithm of \sig : the correlation coefficient, 
r$=-0.73$, has a probability as low as  $\sim3 \times 10^{-6}$ that the data
are randomly extracted from an uncorrelated parent population. The observed and intrinsic (subtracting in quadrature the data uncertainties) spreads are: 1.12 dex and 1.00 dex, respectively.

A correlation between the AGN luminosity and the X-ray variability (without using the line width of a BLR line as a second parameter) of the type

\begin{equation}
{\rm log}{\rm L\over{\rm erg~ s^{-1}}}   =  \alpha '\cdot{\rm log} \sigma^2_{\rm rms}  + \beta ',
\label{eq:relation3}
\end{equation}

\noindent
was already reported \citep[e.g.][and references therein]{ponti12}. If the same sample is used, the linear correlation between logL$_{5100}$ and log\sig  has  a spread of 1.78 dex (instead of 1.12 dex) while the correlation coefficient is -0.36 instead of -0.73  (see Figure \ref{fig_1} and Table \ref{tab:fit}).
It is then evident that the virial product is significantly better correlated with the AGN variability than the luminosity alone.

Slightly better results are obtained if the intrinsic 2-10 keV luminosity, L$_{\rm X}$, instead of the optical luminosity, 
L$_{\rm 5100}$,  is used to compute the virial product. In this case the total and intrinsic spreads are 1.06 dex and 0.93 dex, respectively
(see Figure \ref{fig_1} and Table \ref{tab:fit}).  Also in this case the virial product is better correlated with \sig\ (r=-0.81 and probability $5\times 10^{-10}$) than L$_{\rm X}$ alone is (r=-0.57 and spread 1.36 dex). 

Finally,
if the virial product is computed using L$_{\rm X}$ and Pa$\beta$, the spreads considerably decrease  down to 0.71 dex (total) and 0.56 dex (intrinsic), while the correlation coefficient results to be r$=-0.82$ with a probability of $3\times 10^{-5}$ (see Table \ref{tab:fit} and Figure \ref{fig_1}). The correlation between L$_{\rm X}$ only and \sig\ has instead a less significant coefficient  r$=-0.63$ (probability $4\times 10^{-3}$) and a larger spread of 1.33 dex.

\begin{deluxetable*}{l c c c c l c c  }
\tablecaption{Parameters of the fits}
\tablewidth{0pt}
\tablehead{
\colhead{Variables} & \colhead{$\alpha$} & \colhead{$\beta$} & \colhead{N. Obj} &\colhead{r} &  \colhead{Prob(r)} & \colhead{Spread} & \colhead{Intrinsic Spread} \\
\colhead{}          & \colhead{}           & \colhead{}         & \colhead{}  &  \colhead{}  &  \colhead{}         & \colhead{dex}  & \colhead{dex}  \\
\colhead{(1)} & \colhead{(2)} &\colhead{(3)} &\colhead{(4)} &\colhead{(5)} &\colhead{(6)} &\colhead{(7)} &\colhead{(8) }  
}
\startdata
L$_{5100}$, H$_\beta$       & -1.74$\pm$0.13 & 41.17$\pm$0.29 & 31 & -0.734 & 3$\times 10^{-6}$ & 1.12 & 1.00  \\
L$_{5100}$                          & -1.98$\pm$0.11 & 39.45$\pm$0.20 & 31 & -0.363 & 5$\times 10^{-2}$ & 1.78 & 1.72  \\
\\
L$_{X}$, H$_\beta$       & -1.89$\pm$0.10 & 40.59$\pm$0.23 & 38 & -0.813 & 5$\times 10^{-10}$ & 1.06 & 0.93  \\
L$_{X}$                          & -1.67$\pm$0.07 & 39.90$\pm$0.12 & 38 & -0.570 & 2$\times 10^{-4}$ & 1.36 & 1.32  \\
\\
L$_{X}$, Pa$_\beta$       & -1.21$\pm$0.12 & 41.99$\pm$0.31 & 18 & -0.822 & 3$\times 10^{-5}$& 0.71 & 0.56  \\
L$_{X}$                          & -1.64$\pm$0.09 & 39.58$\pm$0.18 & 18 & -0.634 & 5$\times 10^{-3}$& 1.33 & 1.28  
\enddata
\tablecomments{Column 1: variables used to  compute either the virial product or the luminosity. Columns 2 and 3: best fit parameters of eq. \ref{eq:relation2}. Column 4: number of objects used. Column 5: correlation coefficient. Column 6: probability of the correlation coefficient. Column 7: logarithmic spread of the data on the y axis.
Column 8: intrinsic   logarithmic spread of the data on the y axis.}
\label{tab:fit}
\end{deluxetable*}

\section{Discussion and Conclusions}
\label{Sec:disc}

The above described fits show that, as expected, highly significant relationships exist between the virial products and the AGN X-ray flux variability. These relationships
allow us to predict the AGN 2-10 keV luminosities. The less scattered relation has a spread of 0.6-0.7 dex and is obtained when the Pa$\beta$ line width is used.
 This could be due either because the Pa$\beta$ broad emission line, contrary to H$\beta$, is observed to be practically unblended with other chemical species, or because,  as our analysis is based on a collection of data from public archives, the Pa$\beta$ line widths, 
which comes from the same project \citep{landt08, landt13},  could have therefore been measured in a more homogeneous way.
In this case, it is then probable that new dedicated homogeneous observing programs could obtain even less scattered calibrations; at least for the H$\beta$-based relationships discussed in this work.

In order to use this method to measure the cosmological distances and then the curvature of the Universe,
it is necessary to obtain reliable variability measures, corrected for the cosmological time-dilation, at relevant redshifts. In this respect, the relations based on the H$\beta$ line width measurement are the most promising as can be used even up to redshift $\sim$3 via near infra-red spectroscopic observations (e.g. in the 1-5 $\mu m$ wavelength range with NIRSpec\footnote{See \url{http://www.stsci.edu/jwst/instruments/nirspec}} on the {\it James Webb Space Telescope}). 
Moreover, recent studies by \citet{lanzuisi14} suggest that previous claims of a dependence
on redshift of the AGN X-ray variability should be attributed to selection effects.

\begin{figure}
 \centering
  \includegraphics[width=8.8cm, angle=0]{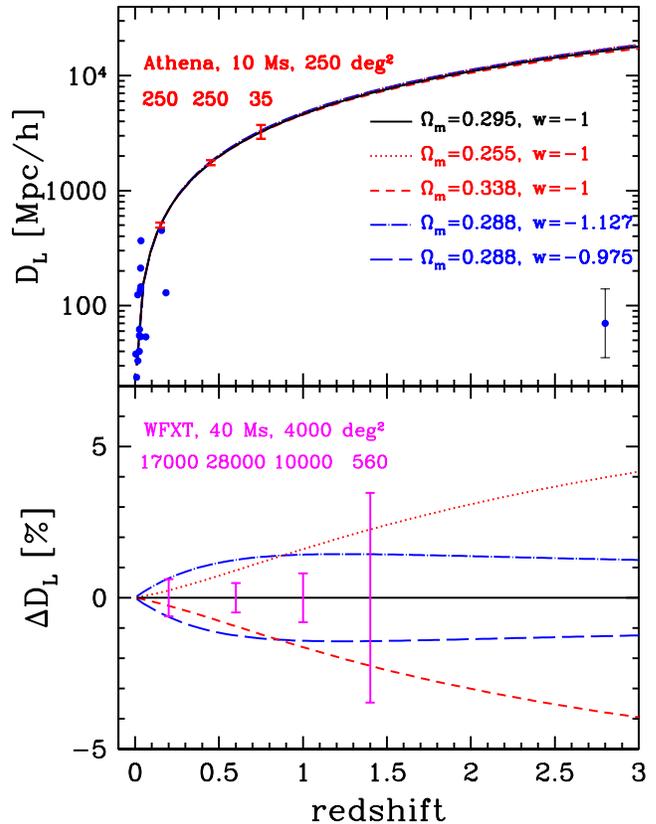}
  \caption{{\it Top.} Luminosity distance as a function of redshift.
The curves represent cosmological models allowed by different datasets described in the text.
Our measures, using L$_X$ and Pa$\beta$, are shown by blue dots. On the lower right corner the  typical uncertainty on a single measurement is shown. Red error bars show the expected uncertainties from a survey carried out with {\it Athena}. The expected number of AGNs in each of the 0.4-wide redshift bins is also shown. {\it Bottom.}  Percent differences of the various cosmological models compared to their respective best fit.
 Magenta error bars show the expected uncertainties from a survey carried out with a future WFXT as described in the text. The expected number of AGNs in each of the 0.3-wide redshift bins is also shown.}
        \label{fig_4}
 \end{figure}

Our AGN-based relations constitute a distance indicator alternative to  SNeIa and
GRB, that can be used to cross-check their distance estimates, revealing potential
unknown sources of systematic errors in their calibration and improve the
constraints on fundamental cosmological parameters including dark energy properties.
To assess cosmological relevance of our distance estimate we compare, in 
Figure \ref{fig_4},  the luminosity distance, D$_L$,  of our estimator (blue dots) with two different sets of cosmological models. The first one
refers to flat $\Lambda$CDM models 
allowed by the Union2.1 compilation of SNeIa \citep{suzuki12}. The black curve represents the best fit, while the red dashed and dotted curves
are the $\pm 1\sigma$ bounds. The corresponding values for $\Omega_{\rm M}$ are indicated in the plot.
The second sets of curves represent
a flat  Dark Energy models with a non-evolving equation of state ($w$CDM), i.e. 
with constant $w$-parameter ($w \equiv p/\rho$), consistent with
both the Planck maps and  galaxy clustering in the BOSS survey \citep{sanchez13}, but
with no reference to SNeIa data. The two blue dot-dashed and dashed curves 
represent $\pm 1\sigma$ bounds with cosmological parameters indicated in the plot.

From a cosmological viewpoint our present application should be considered as a proof
of concept that, however, can be developed by future missions such as the new mission concept 
{\it Athena} \citep[recently proposed to the European Space Agency;][]{nandra13}.
As D$_{\rm L}$ is proportional to the square root of the luminosity,
the 0.7 dex uncertainty on the prediction of the AGN X-ray luminosity corresponds to a 0.35 dex uncertainty on the D$_{\rm L}$ measurement  (see lower right corner in the upper panel of Figure \ref{fig_4})\footnote{The logarithmic uncertainties on D$_{\rm L}$ should be multiplied by a factor 5 to convert them into  distance modulus, $\Delta\rm M$, units.}.  This implies that, if log-normal errors are assumed, variability measures of samples
containing a number  of AGN, N$_{\rm AGN}$, all having similar redshifts, will provide measures of the distance (at that average redshift) with uncertainties of $\sim$0.35/$\sqrt {\rm N_{\rm AGN }}$ dex.
From \citet{vaugh03}, in low signal-to-noise measurement conditions (when the Poissonian noise dominates), the excess variance measurement is larger than the noise when

\begin{equation}
\sigma^2_{\rm rms} > \sqrt{2\over \rm  N}{1 \over {\mu t_o}},
\label{eq_lim}
\end{equation}
where N is the number of $t_o$ long time intervals, and $\mu$ is the average count rate in $ph/s$ units.  As also confirmed by our data, the above formula requires a count rate $\mu \sim 1~ ph/s$ and 80 bins, $t_o= 250$ s long, in order to measure \sig\ larger than $\sim 5\times10^{-4}$ (as mainly observed in this work).
If {\it Athena} will be used, $\mu \sim 1~ ph/s$ corresponds to a 2-10 keV flux of 10$^{-13}$ erg s$^{-1}$ cm$^{-2}$. According to the AGN X-ray luminosity function
\citep{lafranca05, gilli07}, at these fluxes, with a 10 Ms survey covering 250 deg$^2$ with 500 pointings of the Wide Field Instrument ($\sim$0.5 deg$^2$ large field of view), it will be possible to measure \sig\  in a sample of $\sim$250 unabsorbed (N$_{\rm H}<$10$^{21}$ cm$^{-2}$) AGN contained in each of the redshifts bins 0$<z<$0.3 and 0.3$<z<$0.6, and a sample of $\sim$35 AGN in the redshift bin 0.6$<z<$0.9. In this case D$_{\rm L}$ could be measured with a 0.02 dex uncertainty (0.1 mag) at redshifts less than 0.6,  and with a 0.06 dex (0.3 mag) uncertainty in the 0.6$<z<$0.9 bin (red error-bars in Figure \ref{fig_4}). 
With the proposed {\it Athena} survey our estimator will not be competitive with SNeIa. It will, however, provide a
cosmological test independent from SNeIa able to detect possible systematic errors if larger than 0.1 mag in the redshift range $z<0.6$.
A value a factor of $\sim 4$ more precise than the other alternative estimator based on the GRBs \citep{schaefer07}.

In order to significantly exploit at higher redshifts our proposed \sig-based AGN luminosity indicator to constraint the Universe geometry 
a further step is necessary, such as a
dedicated Wide Field X-ray Telescope (WFXT) with an effective collecting area at least three times larger than {\it Athena} and $\sim$2 deg$^2$ large field of view\footnote{\smallskip A similar kind of mission has already been proposed \citep[][see also: \url{http://wfxt.pha.jhu.edu/index.html}]{conconi10}.}. In this case, as an example, with a 40 Ms long program it would be possible to measure D$_{\rm L}$ with less than 0.003 dex (0.015 mag) uncertainties at redshift below 1.2 and an uncertainty of less than 0.02 dex (0.1 mag) in the redshift range $1.2<z<1.6$.
The bottom panel of Figure  \ref{fig_4} illustrates more clearly
the potential of our new estimator. The curves represent
the per cent difference of the luminosity distance models shown in the upper panel with respect to its reference best fit scenario.
From the comparison between the magenta error-bars with the model scatter, we conclude that
our estimator has the prospect to become a cosmological probe even more sensitive than current SNeIa 
if applied to AGN samples as large as that of an hypothetical future survey carried out with a dedicated WFXT as described above.

\acknowledgments 
We thanks Federico Marulli for having helped us in the production of the luminosity distance dependences on redshift as a function
of the cosmological models shown in this work. Paola Marziani, Dario Trevese and Fausto Vagnetti are acknowledged for useful discussions.
Francesca Onori and Federica Ricci are acknowledged for having contributed in the archival
data research, suggestions and carefully reading the manuscript. GP acknowledges support via an EU Marie Curie Intra-European fellowship under contract FP-PEOPLE-2012-IEF-331095. This project was supported by the international mobility program of University Roma Tre.
We acknowledge  funding from PRIN/MIUR-2010 award 2010NHBSBE. 
We would like to thank the referee for helpful comments.

\bibliographystyle{apj}






\end{document}